\documentclass[aps,prb,twocolumn,groupedaddress,showpacs,showkeys,amsmath,amssymb]{revtex4}

\usepackage{amsfonts}
\usepackage{amssymb,amsmath}
\usepackage{graphicx}
\usepackage{dcolumn}
\usepackage{bm}

\begin{document}

\title{Finite-temperature order-disorder phase transition
       in a frustrated  bilayer quantum Heisenberg antiferromagnet
       in strong magnetic fields}

\author{Johannes Richter$^1$,
        Oleg Derzhko$^{1,2,3}$
        and
        Taras Krokhmalskii$^2$}
\affiliation{
\it $^{1}$Institut f\"{u}r Theoretische Physik,
          Universit\"{a}t Magdeburg,
          P.O. Box 4120, D-39016 Magdeburg, Germany\\
\it $^{2}$Institute for Condensed Matter Physics,
          National Academy of Sciences of Ukraine,
          1 Svientsitskii Street, L'viv-11, 79011, Ukraine\\
\it $^{3}$National University ``Lvivska Politechnika'',
          12 S.~Bandera Street, L'viv, 79013, Ukraine}

\date{\today}

\pacs{75.10.Jm, 75.45.+j}

\keywords{bilayer antiferromagnet,
          geometric frustrations,
          high magnetic fields,
          hard-square problem}

\begin{abstract}
We investigate the thermodynamic properties
of the frustrated bilayer quantum Heisenberg antiferromagnet
at low temperatures in the vicinity of the saturation magnetic field.
The low-energy degrees of freedom of the spin model
are mapped onto a hard-square gas on a square lattice.
We use exact diagonalization data for finite spin systems
to check the validity of such a description.
Using a classical Monte Carlo method
we give a quantitative description
of the thermodynamics of the spin model 
at low temperatures around the saturation field.
The main peculiarity of the considered two-dimensional 
Heisenberg antiferromagnet
is related to a phase transition
of the hard-square model on the square lattice, which 
belongs to the two-dimensional Ising model universality class.
It manifests itself 
in a logarithmic (low-)temperature singularity of the specific heat of the 
spin system observed for magnetic fields 
just below the saturation field. 
\end{abstract}

\maketitle


\section{Introduction}
\label{sec1}

The quantum Heisenberg antiferromagnet (HAFM)
on geometrically frustrated lattices
has attracted much attention during last years.\cite{LNP04,diep05}
Besides intriguing quantum ground-state phases at zero magnetic field 
those systems often show unconventional properties 
in finite magnetic fields like plateaus and jumps in the magnetization curve, 
see e.g. Ref. \onlinecite{ho04}. 
The recent finding
that a wide class of geometrically frustrated quantum spin antiferromagnets
(including the kagom\'{e}, checkerboard and pyrochlore lattices)
has quite simple ground states
in the vicinity of the saturation field,\cite{lm01} 
namely independent localized-magnon states,
has further stimulated studies 
of the corresponding frustrated quantum antiferromagnets
at high magnetic fields.\cite{prl04,cool04,zhi_tsu,entro04,euro06}
In particular, the 
low-temperature high-field thermodynamics
of various one- and two-dimensional frustrated quantum antiferromagnets
which support localized-magnon states, 
can be discussed from a quite universal point of view  
by mapping the low-energy degrees of freedom of the quantum HAFM 
onto lattice gases of hard-core objects.\cite{cool04,zhi_tsu,entro04,euro06,zhi_tsu_unp}
For instance,  
the  kagom\'{e} (checkerboard) HAFM in the vicinity of the saturation field 
can be mapped onto a gas of hard hexagons (squares) 
on a triangular (square) lattice.\cite{zhi_tsu,entro04,euro06,zhi_tsu_unp}
The exactly soluble hard-hexagon model 
exhibits an order-disorder second-order phase transition.\cite{bax}
The hard-core lattice-gas model corresponding to the checkerboard HAFM 
consists of large hard squares on the square lattice 
with edge vectors 
$\vec a_1=(2,0)$ and $\vec a_2=(0,2)$ 
(i.e. there is a nearest-neighbor and next-nearest-neighbor exclusion).
For the latter model no exact solution is available, 
but most likely there is also an order-disorder phase transition.\cite{large_sq}
The existence of a phase transition in the hard-hexagon (large-hard-square) model  
would imply a corresponding finite-temperature transition of the corresponding spin model
near saturation provided the low-temperature physics 
is correctly described by the hard-core lattice-gas model.  
However, 
at the present state of the investigations 
no conclusive statements 
for the kagom\'{e} and checkerboard antiferromagnets are available, 
since both models admit additional degenerate eigenstates 
not described by the hard-hexagon/large-hard-square model.\cite{ho_priv,hjs06} 
Furthermore for these spin models  precise statements 
on the gap bet\-ween the localized-magnon ground states and the excitations 
are not available. 
Therefore, 
the effect of additional ground states and the excited states 
on the low-temperature thermodynamics 
remains unclear.

The motivation for the present paper 
is to find and discuss 
another two-dimensional frustrated quantum HAFM,
for which a hard-core lattice gas
completely covers all low-energy states of the spin model
in the vicinity of the saturation field 
and 
where all excitations are separated by a finite energy gap.
For such a spin model 
one can expect that an order-disorder phase transition 
inherent in the hard-core lattice-gas model  
can be observed 
as a finite-temperature phase transition 
in the spin model.  
It might be worth to note 
that such a phase transition of course does not contradict 
the Mermin-Wagner theorem\cite{mermin} 
that forbids 
magnetic long-range order 
(breaking the rotational symmetry) 
for the two-dimensional Heisenberg model 
at any non-zero temperature and at zero field. 

A spin model which satisfies these requirements 
is a frustrated bilayer quantum HAFM.
The investigation of the bilayer quantum HAFM
was  initially motivated by bilayer high-$T_c$ superconductors\cite{scal94}
and has been continued till present time, 
see e.g. Ref. \onlinecite{wang06} and references therein.
Below we will illustrate 
that the corresponding hard-core lattice-gas model 
is a model of hard squares on a square lattice, 
however, 
in difference to the checkerboard lattice with smaller hard squares 
with edge vectors 
$\vec a_1=(1,1)$ and $\vec a_2=(-1,1)$ 
(i.e. there is a nearest-neighbor exclusion, only, cf. Fig.~\ref{fig01}).
This hard-square model   
exhibits an order-disorder phase transition.\cite{bax,bax80,bax99}
In the context of different universality classes discussed in Ref. \onlinecite{euro06}
the frustrated bilayer quantum HAFM 
is the first example of a spin system 
which belongs to the universality class of (small) hard squares.

The paper is organized as follows.
First, 
we specify the frustrated bilayer model 
and illustrated the corresponding localized-magnon states
(Sec.~\ref{sec2}).
Then 
in Sec.~\ref{sec3} 
we calculate the contribution of the independent localized-magnon states 
to the thermodynamic quantities
using the hard-square model.
We compare our results with exact diagonalization data 
for finite spin-1/2 Heisenberg systems of up to $N=32$ sites.
Finally,
in Sec.~\ref{sec4} 
we report the low-temperature high-field thermodynamic quantities
obtained on the basis of Monte Carlo simulations for hard squares
focusing on the (low-)temperature dependence of the specific heat 
in the vicinity of the saturation field.

\section{The frustrated bilayer antiferromagnet and independent localized magnons}
\label{sec2}

\begin{figure}
\begin{center}
\includegraphics[clip=on,width=80mm,angle=0]{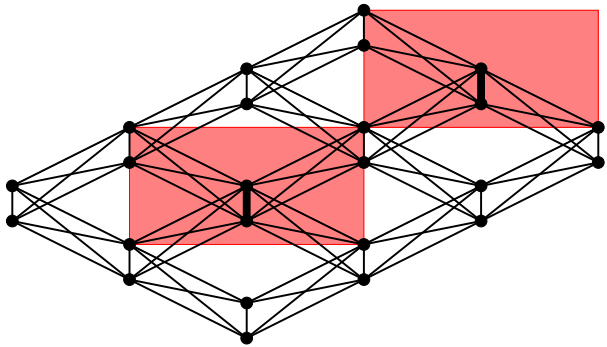}
\caption
{(Color online) 
The frustrated bilayer antiferromagnet (black lines).
The vertical bonds have the strength $J_2$
whereas all other bonds have the strength $J_1$.
The trapping cells (vertical bonds) occupied by localized magnons
are shown by fat lines.
The auxiliary square lattice is a simple square lattice 
filled by hard squares 
(indicated as red squares)
which correspond to localized magnons.}
\label{fig01}
\end{center}
\end{figure}

To be specific,
we consider the nearest-neighbor Heisenberg antiferromagnet
in an external magnetic field
on the lattice shown in Fig.~\ref{fig01}.
This lattice may be viewed 
as a two-dimensional version 
of the frustrated two-leg ladder considered in Refs.~\onlinecite{gelfand,mila,ho00}
(for some similar models see Ref. \onlinecite{chen}).
The Heisenberg Hamiltonian
of $N$ quantum spins of length $s$
reads
\begin{eqnarray}
\label{01}
H=
\sum_{(nm)}J_{nm}\left (\frac{1}{2}\left(s_n^+s_m^-+s_n^-s_m^+\right)
+\Delta s_n^zs_m^z\right)
-hS^z.
\end{eqnarray}
Here
the sum runs over the bonds (edges)
which connect the neighboring sites (vertices)
on the spin lattice shown in Fig.~\ref{fig01}.
$J_{nm}>0$ are the antiferromagnetic exchange constants
between the sites $n$ and $m$
which take two values,
namely,
$J_2$ for the vertical bonds
and
$J_1$ for all other bonds.
$\Delta\ge 0$ is the exchange interaction anisotropy parameter,
$h$ is the external magnetic field,
and
$S^z=\sum_ns_n^z$ is the $z$-component of the total spin.
In our exact diagonalization studies reported below 
we will focus 
on $s=1/2$ and $\Delta=1$.

We note that the Hamiltonian (\ref{01}) commutes with the operator $S^z$
and hence 
we may consider the subspaces of its eigenstates with different values of $S^z$ separately.
Evidently, 
the fully polarized state
$\vert 0 \rangle =\vert s,\ldots,s\rangle$
is the eigenstate of the Hamiltonian (\ref{01}) with $S^z=Ns$ 
and can be considered 
as the vacuum state with respect to the number of excited magnons.  
This state is the ground state for high magnetic fields.

Consider next the one-magnon subspace with $S^z=Ns-1$.
The one-particle energy is given by
$\Lambda_{{\bf{k}}}^{(1)}
=-s\left(J_2+\Delta\left(8J_1+J_2\right)\right)+h$
and
$\Lambda_{{\bf{k}}}^{(2)}
=s\left(4J_1\left(\cos k_x+\cos k_y\right)
+J_2-\Delta\left(8J_1+J_2\right)\right)+h$
(here
$k_{\alpha}=2\pi n_{\alpha}/M_{\alpha}$,
$n_{\alpha}=1,2,\ldots,M_\alpha$,
$\alpha=x,y$, 
$M_xM_y=N/2$).
Obviously,
the excitation branch
$\Lambda_{{\bf{k}}}^{(1)}$
is dispersionless
and it becomes the lower one when
$J_2>4J_1$.
Throughout this paper we assume $J_2\ge 4J_1$.
Then the saturation field is given by
$h_1=s\left(J_2+\Delta\left(8J_1+J_2\right)\right)$.

The $N/2$ dispersionless one-magnon 
excitations can be written as localized excitations
on the $N/2$  vertical bonds, 
i.e.
$\vert 1 \rangle=\vert {\rm{lm}}\rangle_v \vert s,\ldots,s\rangle_e$
is an eigenstate of (\ref{01}) in the subspace with $S^z=Ns-1$
with the zero-field  eigenvalue
$E_{{\rm{FM}}}-\epsilon_1$, where
$E_{{\rm{FM}}}=4N\Delta s^2J_1+N\Delta s^2J_2/2$
and
$\epsilon_1=s\left(J_2+\Delta\left(8J_1+J_2\right)\right)$.
In 
$\vert 1 \rangle$ the first part is 
the localized one-magnon excitation on 
the vertical bond number $v$, i.e.    
$\vert {\rm{lm}}\rangle_v= 2^{-1/2}\left( 
\vert {s},{s-1}\rangle-\vert {s-1},{s}\rangle\right )_v$
and the second part 
$\vert s,\ldots,s\rangle_e$ 
is the fully polarized environment.

We pass to the subspaces with $S^z=Ns-2,Ns-3,\ldots,Ns-n_{\max}$,
where $n_{\max}=N/4$.
We can easily construct many-particle states in these subspaces
using the localized-magnon states.
Explicitly 
the wave function of $n$ independent localized magnons has the form
\begin{eqnarray}
\label{02}
\vert n \rangle
=
\vert {\rm{lm}}\rangle_{v_1}
\ldots 
\vert {\rm{lm}}\rangle_{v_i} 
\ldots
\vert {\rm{lm}}\rangle_{v_n}
\vert s,\ldots, s\rangle_e.
\end{eqnarray}
It is important to note
that any two vertical bonds  $v_i$ and $v_j$ 
 in Eq. (\ref{02})
where localized magnons live
are not allowed  to be direct neighbors.
The energy of the $n$ independent localized-magnon state (\ref{02})
in zero field $h=0$ is
$E_n=E_{{\rm{FM}}}-n\epsilon_1$.
Since $n$ independent localized magnons can be put on the bilayer in many ways 
the eigenstates (\ref{02}) are highly degenerate.
We denote this degeneracy by $g_N(n)$,
that is the number of ways 
to put $n$ hard squares on a lattice of ${\cal{N}}=N/2$ sites
(see Fig.~\ref{fig01}).
According to Refs.~\onlinecite{lm01,hjs02}
the independent localized-magnon states (\ref{02})
are the states with the lowest energy in the corresponding subspaces.
Moreover,
they are linearly independent
(orthogonal type in the nomenclature of Ref. \onlinecite{hjs06})
and form an orthogonal basis in each subspace.\cite{hjs06}
Due to their linear independence 
they all contribute to the partition function of the spin system.

In the presence of an external field 
the eigenstates (\ref{02}) have the energy
$E_n(h)=E_{{\rm{FM}}}-hsN-n(\epsilon_1-h)$.
At the saturation field, 
$h=h_1=\epsilon_1$, 
all they are ground states
and the ground-state energy $E_n(h_1)$ does not depend on $n$.
As a result
the ground-state magnetization curve exhibits a jump at the saturation field.
This jump is accompanied by a preceding wide plateau, 
where the width of this plateau can be obtained 
following the arguments given in Ref. \onlinecite{ho00} 
and from finite-size data.
We find for $s=1/2$, $\Delta=1$ a plateau width of $h_1-h_2=4J_1$.
This plateau belongs to the two-fold degenerate ground state with 
maximum density $n_{\max}=N/4$ of localized magnons, the so-called 
magnon crystal,\cite{lm01} where all localized magnons occupy only one of 
the two sublattices  of the underlying square lattice.
  
Furthermore,
the degeneracy of independent localized-magnon states at the saturation field 
${\cal{W}}=\sum_{n=0}^{n_{\max}}g_N(n)$
grows exponentially with the system size $N$
that implies a nonzero ground-state residual entropy 
${\cal{S}}=k\lim_{N\to\infty}(\ln{\cal{W}}/N) 
= 0.2037\ldots k$ 
(see Ref. \onlinecite{bax99} and also below).
Due to their high degeneracy the independent localized-magnon 
states are also dominating the thermodynamic properties   
at low temperatures for magnetic fields around the saturation field
as we will dicuss in detail below.

\section{Hard-square model}
\label{sec3}

We want to calculate 
the contribution of the independent localized magnons
to the canonical partition function of the spin system,
\begin{eqnarray}
\label{03}
Z_{{\rm{lm}}}(T,h,N)
=\sum_{n=0}^{n_{\max}}g_N(n)\exp\left(-\frac{E_n(h)}{kT}\right)
\nonumber\\
=\exp\left(-\frac{E_{{\rm{FM}}}-hsN}{kT}\right)
\sum_{n=0}^{n_{\max}}g_N(n)\exp\left(\frac{\mu}{kT}n\right),
\end{eqnarray}
where $\mu=\epsilon_1-h=h_1-h$.
It is apparent 
that $g_N(n)$ is the canonical partition function $Z(n,{\cal{N}})$
of $n$ hard squares on a square lattice of ${\cal{N}}=N/2$ sites,
whereas 
$\Xi(T,\mu,{\cal{N}})
=\sum_{n=0}^{n_{\max}}g_N(n)\exp\left(\mu n/kT\right)$
is the grand canonical partition function 
of hard squares on a square lattice of ${\cal{N}}=N/2$ sites
and $\mu$ is the chemical potential of the hard squares.
As a result
we arrive at the basic relation between the localized-magnon contribution
to the canonical partition function of the spin model
and the grand canonical partition function of hard-square model,
\begin{eqnarray}
\label{04}
Z_{{\rm{lm}}}(T,h,N)
=\exp\left(-\frac{E_{{\rm{FM}}}-hsN}{kT}\right)\Xi(T,\mu,{\cal{N}}).
\end{eqnarray}
Eq. (\ref{04}) yields the Helmholtz free energy of the spin system 
$F_{{\rm{lm}}}(T,h,N)=-kT\ln Z_{{\rm{lm}}}(T,h,N)$,
whereas
the entropy $S$, 
the specific heat $C$ 
and the magnetization $M=\langle S^z\rangle$
are given by the usual formulas,
$S_{{\rm{lm}}}(T,h,N)=-\partial F_{{\rm{lm}}}(T,h,N)/\partial T$,
$C_{{\rm{lm}}}(T,h,N)=T\partial S_{{\rm{lm}}}(T,h,N)/\partial T$,
$M_{{\rm{lm}}}(T,h,N)=sN-kT\partial \ln\Xi(T,\mu,{\cal{N}})/\partial\mu$.
\par
We use exact diagonalization data to check this picture
for $s=1/2$ isotropic Heisenberg system (\ref{01}) 
of $N=16$ and $N=20$ sites 
(full diagonalization)
and $N=32$
(only in the subspaces with $S^z=16,\ldots,11$)
imposing periodic boundary conditions.
We fix the energy scale  by putting  
$J_1=1$. 
For the  vertical exchange bonds we  consider $J_2=4,\;5,\;10$. 

First we compare the  degeneracies $g_N(n)$ of the localized $n$-magnon states 
calculated for spin systems of sizes $N=16,20,32$ 
with the corresponding values $Z(n,N/2)$ 
of the hard-square model. 
As expected 
we find complete agreement between both models for $J_2>4J_1$.
As an example 
we give here the numbers for $N=20$: 
$g_{20}(n)=1,\;10,\;25,\;20,\;10,\;2$ 
for $n=0,\;1,\;2,\;3,\;4,\;5$. 
For $J_2=4J_1$ 
the spin system has one extra state for $n=1$, 
i.e. in the one-magnon sector, 
since the dispersive mode
$\Lambda_{{\bf{k}}}^{(2)}$ 
and the dispersionless mode 
$\Lambda_{{\bf{k}}}^{(1)}$ 
are degenerate at ${\bf{k}}=(\pi,\pi)$.

To estimate the relevance of excited states, 
not described by the localized-magnon scenario, 
we have determined 
the thermodynamically relevant energy separation $\Delta_{\rm{DOS}}$   
between the localized-magnon states 
and the other states of the spin system
by calculating the integrated low-energy density of states at saturation field. 
We define $\Delta_{\rm{DOS}}$ 
as that energy value above the localized-magnon ground-state energy,
where the contribution of the higher-energy states to the integrated density of states 
becomes as large 
as the contribution of the localized-magnon states.
For both values
$J_2=5$ and $J_2=10$
we find 
$\Delta_{\rm{DOS}}\approx 1$ 
independent of the size of the system $N$. 
We can expect 
that the contribution of the localized-magnon states to the partition function 
is dominating for temperatures $kT$ 
significantly smaller than $\Delta_{\rm{DOS}}$.  

\begin{figure}
\begin{center}
\includegraphics[clip=on,width=75mm,angle=0]{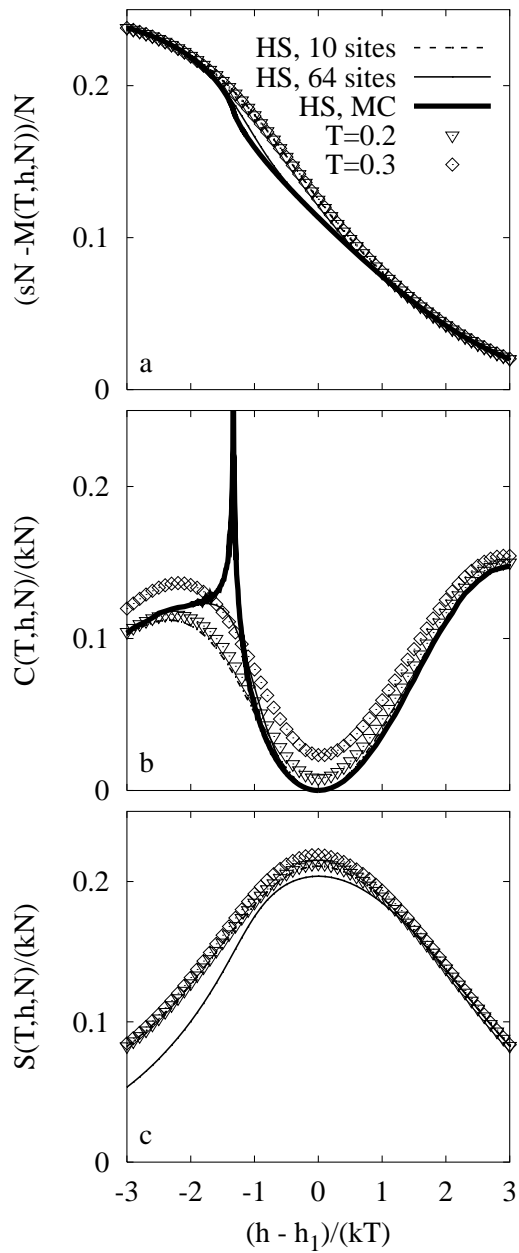}
\caption
{The magnetization $(sN-M(T,h,N))/N$ (a),
the specific heat $C(T,h,N)/kN$ (b)
and the entropy $S(T,h,N)/kN$ (c)
as functions of $(h-h_1)/kT$.
Symbols correspond to the exact diagonalization data
($N=20$, $J_2=5$, $kT=0.2$ (triangles), $kT=0.3$ (diamonds)).
Thin lines correspond to the results for a small hard-square system
${\cal{N}}=10$ (thin dotted lines)
and 
${\cal{N}}=64$ (thin solid lines).
The exact diagonalization data 
for the temperatures up to $kT=0.1$ 
coincide with the corresponding data 
for the ${\cal{N}}=10$ hard-square system.
We also show by thick lines 
the Monte Carlo simulation results 
for the magnetization and the specific heat 
for a hard-square system of sizes up to $800\times 800$ sites.}
\label{fig02}
\end{center}
\end{figure}

In Fig.~\ref{fig02}
we present our results 
for the magnetization, the specific heat and the entropy 
(panels from top to bottom)
for the spin system of size $N=20$ 
(triangles  and diamonds).
We compare those data 
with the corresponding data 
taking into account only the contribution of independent localized-magnon states, 
described by the finite hard-square model of ${\cal N}=10$ sites
(thin dotted lines).
Up to $kT \approx 0.1$ both data sets coincide, 
demonstrating that the hard-square description perfectly works at low temperatures. 
But we observe good agreement also for higher temperatures up to $kT \approx 0.3$, 
indicating that the independent localized magnons 
still dominate the thermodynamic quantities.
Further increasing $kT$, 
the high-energy states more and more contribute to the partition function 
and the hard-square description loses its validity.
Note that an identical statement can be made for $N=16$ (${\cal N}=8$). 

\section{Low-temperature thermodynamics around the saturation field}
\label{sec4}

\begin{figure}
\begin{center}
\includegraphics[clip=on,width=75mm,angle=0]{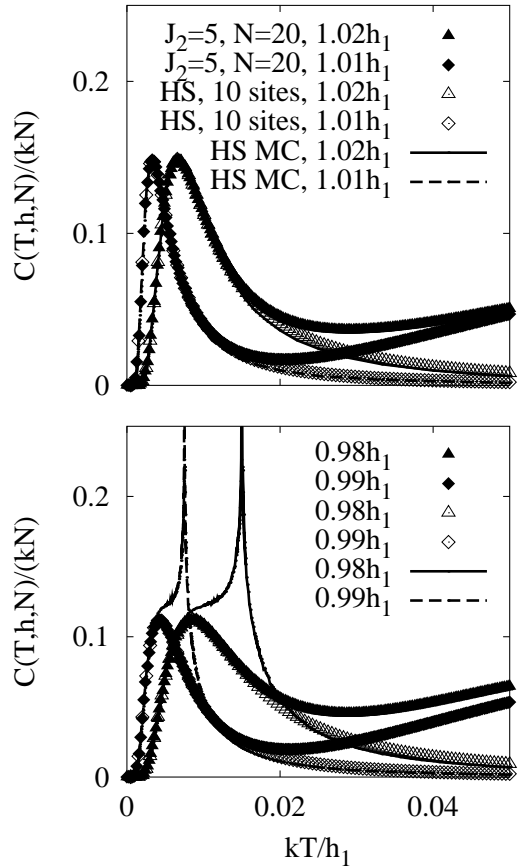}
\caption
{The temperature dependence of the specific heat $C(T,h,N)/kN$
for $h=1.01h_1$ (broken lines, diamonds) 
and $h=1.02h_1$ (solid lines, triangles)
(upper panel)
and 
for $h=0.99h_1$ (broken lines, diamonds) 
and $h=0.98h_1$ (solid lines, triangles) 
(lower panel).
The results of Monte Carlo simulations 
for a hard-square system of sizes up to $800\times 800$ sites
are shown by lines.
We also report 
the corresponding results 
for the hard-square system with ${\cal{N}}=10$ (open symbols)
and 
the exact diagonalization data for the finite spin system with $N=20$, $J_2=5$ (filled symbols).}
\label{fig03}
\end{center}
\end{figure}

Now we discuss 
the low-temperature high-field thermodynamics 
of the frustrated bilayer quantum HAFM 
using the results for the hard-square model. 
From Fig.~\ref{fig02}b it is obvious, that  
fixing the temperature to a sufficiently low value  
the spin system can be driven through a phase transition by 
increasing the magnetic field $h$ towards the saturation field $h_1$. 
On the other hand, 
we can 
fix the magnetic field slightly below the saturation field 
and vary the temperature. 
Then the phase transition is driven by the temperature  and the specific 
heat exhibits a singularity at a critical temperature $T_c(h)$, 
see Fig.~\ref{fig03}, 
where we show the results of a Monte Carlo simulation  
for large hard-square systems
(periodic cells with up to $800\times 800$ sites 
and $3\cdot 10^6$ steps) 
for the specific heat for two values of the 
magnetic field.
The data clearly indicate a  phase transition 
which occurs in the hard-square model at $z_c = 3.7962\ldots$, 
i.e. at $((h_1-h)/kT)_c=\ln z_c\approx 1.3340$ which 
yields $kT_c(h) \approx (h_1-h)/1.3340$. 
The corresponding order parameter 
is the difference of the density
(of hard squares or localized magnons 
in hard-square or spin language, respectively) 
on the A- and B-sublattices of the 
underlying square lattice.\cite{bax}
For $((h_1-h)/kT)_c<\ln z_c$ 
(i.e. for $T>T_c(h)$, $h<h_1$)
both sublattices are equally occupied, 
but for $((h_1-h)/kT)_c>\ln z_c$ 
(i.e. for $T<T_c(h)$, $h<h_1$)
one of two sublattices is more occupied than the other. Therefore in the ordered 
phase the translational symmetry of the spin (hard-square) 
system is broken.  
Finally, at $T=0$ only one sublattice is occupied and the other is empty 
and the ground state of the spin system is a magnon-crystal state, 
see Sec.~\ref{sec2}.  

We can estimate the critical temperature 
for a fixed deviation of the field from the saturation value
using the above given expression for $kT_c(h)$.
For $1-h/h_1=0.02$ we find $kT_c/h_1\approx 0.0150$,
whereas
for $1-h/h_1=0.01$ we have $kT_c/h_1\approx 0.0075$
and for the spin system with $J_2=5$ we find
for $h=8.91J_1$ ($h=8.82J_1$)
$kT_c\approx 0.0675 J_1$ ($kT_c\approx 0.1349 J_1$).
Such temperatures are within a temperature range 
where the hard-square description for finite systems works {\em perfectly} well 
(see Figs.~\ref{fig02},~\ref{fig03}). However, 
one may expect that the scenario of the phase transition may hold 
also at temperatures, 
for which the localized-magnon states are still dominant 
but also higher-energy  states of the spin system not 
described by the hard-square model contribute to the partition function.

We mention that the hard-square model belongs to the two-dimensional Ising model 
universality class\cite{racz,bax,guo} 
with the critical exponents $\beta =1/8$ for the order parameter 
and $\alpha=0$ for the specific heat,
i.e. 
the specific heat shows a logarithmic singularity at the critical point.
Note 
that this universality class is different  from the one of the hard-hexagon model.\cite{bax}
Thus,
the low-temperature peak (singularity) 
in the temperature dependence of the specific heat 
in the vicinity of the saturation field  
is a spectacular sign 
of highly degenerate independent localized-magnon states 
of the frustrated bilayer quantum HAFM.
Their ordering leads to the hard-square type peculiarity
just below the saturation field.

Another thermodynamic quantity of interest 
is the curve of constant entropy as a function of magnetic field and temperature.
Since hard-square description implies the dependence of the entropy only on $(h_1-h)/kT$
this curve is similar to an ideal paramagnet.
Therefore, the considered spin system is expected 
to exhibit a large magnetocaloric effect 
in the vicinity of the saturation field.\cite{cool04,zhi_tsu,euro06}

Finally we note,
that the effect of the localized magnons is a pure quantum effect
which disappears 
as $s$ increasing approaches the classical limit $s\to\infty$.\cite{lm01}

\section*{Acknowledgments}

The numerical calculations were performed using 
J.~Schulenburg's {\it spinpack}.
The present study was supported by the DFG
(Project No. 436 UKR 17/13/05).
O.~D. acknowledges the kind hospitality of the Magdeburg University
in the summer of 2006.

\end{document}